\documentclass[letterpaper, 10 pt, conference]{ieeeconf} 
\IEEEoverridecommandlockouts 
\usepackage{cite}
\usepackage{amsmath}
\usepackage{graphicx}
\usepackage{algorithmic}
\usepackage{array}
\usepackage{interval}
\usepackage{amssymb,amsfonts,mathrsfs,mathtools}
\usepackage{color}
\usepackage{booktabs,makecell}
\usepackage{pict2e}
\usepackage{flushend}
\makeatletter
 \let\NAT@parse\undefined
 \makeatother
\usepackage{hyperref}
\hypersetup{colorlinks = true,
 linkcolor = red,
 urlcolor = black,
 citecolor = blue,
 anchorcolor = black}
\usepackage{textcomp}
 \usepackage{tikz}

\def\gof{\mbox{${\boldsymbol{P}}\,\#\,\boldsymbol{C}$}}
\newcommand\projtp{\boldsymbol{\Pi}_{\mathscr{G}_e(\sysp)\parallel\mathscr{G}_e^{\dagger}(\sysc)}}

\newcommand\bbkt[1]{\left\{#1\right\}}
\newcommand\sbkt[1]{\left[#1\right]}
\newcommand\rbkt[1]{\left(#1\right)}
\newcommand\ininf[2]{\langle #1\,, #2 \rangle}

\newcommand\lt{\mathcal{L}_2}
\newcommand\ltp{\mathcal{L}_{2}}
\newcommand\ltep{\mathcal{L}_{2e}}
\newcommand\ccp{{\mathbb{C}}_{\geq 0}}
\newcommand\sysp{\boldsymbol{P}}
\newcommand\sysg{\boldsymbol{G}}
\newcommand\srg{\mathrm{SRG}}
\newcommand\gra{\mathscr{G}}
\newcommand\sg{\mathrm{SG}} 
\newcommand\sysi{\boldsymbol{I}}
\newcommand\sysc{\boldsymbol{C}}
\newcommand\sysh{\boldsymbol{H}}
\newcommand\sysf{\boldsymbol{F}}
\newcommand\trucT{\boldsymbol{\Gamma}_T}
\newcommand\rn{{\mathbb{R}}^n}
\newcommand\RI{{\mathrm{I}}}
\newcommand\abs[1]{\left|#1\right|}
\newcommand\rep{{\rm Re}}

\newcommand\imp{{\rm Im}}
\newcommand\dom{{\rm dom}}
\newcommand{\norm}[1]{\left\lVert#1\right\rVert}
\newcommand{\inorm}[1]{\left\lVert#1\right\rVert_{\mathrm{I}}}

\newcommand{\tbt}[4]{\begin{bmatrix}#1&#2\\#3&#4\end{bmatrix}}

\newcommand{\obt}[2]{\begin{bmatrix}#1&#2\end{bmatrix}}
\newcommand{\tbo}[2]{\begin{bmatrix}#1\\#2\end{bmatrix}}

\newcommand{\stbo}[2]{\left[\begin{smallmatrix} #1\\#2\end{smallmatrix}\right]}

\newcommand{\bi}{\begin{itemize}}\newcommand{\ei}{\end{itemize}}
\newcommand{\be}{\begin{equation}}\newcommand{\ee}{\end{equation}}
\newcommand{\bex}{\begin{equation*}}\newcommand{\eex}{\end{equation*}}
\newcommand{\bax}{\begin{align*}}\newcommand{\eax}{\end{align*}}
\newcommand{\bc}{\begin{center}}\newcommand{\ec}{\end{center}}
\def\fpc{\mbox{$\boldsymbol{F}_{\sysp\# \sysc}$}}

\newtheorem{theorem}{Theorem}

\newtheorem{definition}{Definition}
\newtheorem{lemma}{Lemma}

\newtheorem{example}{Example}
\newtheorem{assumption}{Assumption}

\intervalconfig{soft open fences}

\begin{document}
\title{Scaled Relative Graph Separation for Unbounded Nonlinear Systems} 
\author{Chao~Chen, Sei~Zhen~Khong and~Rodolphe~Sepulchre,~\IEEEmembership{Fellow,~IEEE}
\thanks{The research leading to these results has received funding from the
European Research Council under the Advanced ERC Grant Agreement SpikyControl n. 101054323. The
work of Sei Zhen Khong was supported by the National Science and Technology Council of Taiwan under grants 113-2222-E-110-002-MY3, 114-2622-8-110-003, 115-2218-E-007-003, and 115-2221-E-110-058-MY2.} 
\thanks{Chao~Chen is with the Department of Electrical and Electronic Engineering, University of Manchester, Manchester M13 9PL, United Kingdom (e-mail: chao.chen@manchester.ac.uk).}
\thanks{Sei~Zhen~Khong is with the Department of Electrical Engineering, National Sun Yat-sen University, Kaohsiung 80424, Taiwan (e-mail: szkhong@mail.nsysu.edu.tw).} 
\thanks{Rodolphe~Sepulchre is with the STADIUS center of the Department of Electrical Engineering (ESAT), KU Leuven, Leuven 3001, Belgium, and the Department of Engineering, University of Cambridge, Cambridge CB2 1PZ, United Kingdom. (e-mail: rodolphe.sepulchre@kuleuven.be).}
}
\maketitle

\begin{abstract}
We investigate stability analysis of one-channel nonlinear feedback systems based on the recent notion of scaled relative graph (SRG). Our focus is on the SRG separation in the complex plane for feedback stability analysis that involves possibly unbounded open-loop nonlinear systems. Subject to a technical assumption, a homotopy-based SRG separation theorem is developed as the main result, where the incremental directed gap is utilized as a key instrument to quantify robust feedback stability along the homotopy path. Doing so enables us to incorporate commonly used integrators as unbounded systems in the SRG-based framework. Specializations of the main result to linear systems, along with illustrative examples, are also provided to demonstrate its effectiveness. 
\end{abstract}

\IEEEpeerreviewmaketitle

\section{Introduction}
The scaled relative graph (SRG), recently proposed in \cite{Ryu:21, Chaffey:21j, Pates:21}, is a novel graphical tool tailored for operator-theoretic nonlinear systems analysis. The SRG of a nonlinear system translates the \emph{incremental gain} and \emph{incremental phase} information of the system into a collection of complex scalars, largely mimicking the classical Nyquist plot for single-input single-output (SISO) linear time-invariant (LTI) systems. By doing so, the stability of a one-channel nonlinear feedback system can be verified through complex-plane separation of the SRGs of the open-loop systems \cite{Chaffey:21j}, very much akin to the Nyquist stability criterion. Such a result can be viewed as a graphical statement of classical topological graph separation \cite{Teel:96, Georgiou:97, Teel:11, Carrasco:19}. 

The SRG has attracted research interest in systems and control, including \cite{VanDenEijnden:24, VanDenEijnden:25, Chen:25_dominance, Chen:25SRG, Chaffey:24, Baron:25, Quan:24, deGroot:25_dissipativity, Krebbekx:25_SISO, Krebbekx:25_nonlinear, Krebbekx:25computing, Nauta:25, Nauta:26computing, Yang:25}, to name a few. Broadly speaking, there are two main research directions in the current SRG-based theory. One concerns the computational aspects of SRGs for special classes of nonlinear or linear systems, e.g. see \cite{Pates:21, deGroot:25_dissipativity, Nauta:25, Krebbekx:25computing, Nauta:26computing}, since constructing SRGs for general systems is challenging. The other focuses on developing specializations and generalizations of the SRG separation theorem for feedback stability analysis of different types of systems, e.g., see \cite{Chaffey:24, VanDenEijnden:24, VanDenEijnden:25, Chen:25SRG, Chen:25_dominance, Krebbekx:25_SISO, Krebbekx:25_nonlinear, Baron:25, Yang:25, vandenEijnden:26SRG}. In particular, for LTI systems only, the frequency-wise notion of SRG \cite{Chen:25_dominance, Baron:25} is proposed to provide less conservative feedback stability conditions. 

This paper aims to advance the second research direction, particularly focusing on the SRG separation when \emph{unbounded} open-loop systems are considered. Almost all existing SRG separation results, except for the hard SRG \cite{Chen:25SRG} and the SISO extended SRG \cite{Krebbekx:25_SISO}, are developed based on a simple homotopy-based characterization \cite{Megretski:97} of a feedback interconnection of $\tau \sysp$ and $\sysc$, parameterized by $\tau \in \interval{0}{1}$. Such a characterization, however, applies \emph{only} to bounded open-loop systems, which is an unnecessary limitation of the present theory. It would be desirable to extend the theory to enable, for instance, integrators in the feedback analysis. 

This paper is devoted to exploring the SRG separation for one-channel feedback stability analysis with possibly unbounded systems in a direct fashion. Under a technical assumption, the main result is a general homotopy-based SRG separation condition, characterized by the \emph{incremental directed gap} \cite{Georgiou:95}, allowing one of the open-loop systems to be unbounded. The incremental directed gap plays a key role in the development, through which one-channel robust feedback stability is obtained as a by-product, inspired by its non-incremental counterpart studied in nonlinear robust control \cite{Georgiou:97, Georgiou:95, Rantzer:97, Khong:21}. The main result admits a linear specialization, whose usefulness is illustrated by an example. Notably, the hard SRG analysis \cite{Chen:25SRG} requiring the extended $\mathcal{L}_2$ space, the frequency-wise SRG analysis for LTI systems \cite{Chen:25_dominance}, and the extended-SRG method for SISO LTI systems \cite{Krebbekx:25_SISO} also apply to feedback systems involving unbounded systems of different types. By contrast, our treatment of unboundedness in the $\mathcal{L}_2$ space may provide a more direct route to the SRG analysis of nonlinear feedback systems. 

The rest of this paper is organized as follows. Preliminaries on signals, systems, the incremental directed gap, and feedback stability are presented in Section~\ref{sec:preliminaries}. We provide the SRG definition for possibly unbounded systems in Section~\ref{sec: srg} and highlight our motivation using an existing SRG separation theorem. In Section~\ref{sec: main}, the main result for feedback stability analysis is proposed: a novel homotopy-based SRG separation theorem. Finally, Section~\ref{sec:conclusion} concludes this paper.

\section{Notation and Preliminaries}\label{sec:preliminaries}

\emph{Notation:} Let $\mathbb{R}$ be the field of real numbers, and $\mathbb{R}^n$ be the linear space of $n$-dimensional vectors over $\mathbb{R}$. Let $\bar{\mathbb{C}}\coloneqq \mathbb{C}\cup \{\infty\}$, ${\mathbb{C}}_{\geq 0}$ be the extended complex right half-plane, and $\mathbb{R}_{\geq 0}$ be set of the nonnegative real axis. For $x, y\in \mathbb{R}^n$, denote $\ininf{x}{y}\coloneqq x^\top y$ and $\abs{x}\coloneqq \sqrt{\ininf{x}{x}}$ as the Euclidean inner product and norm, respectively. Let $I$ denote the identity matrix of appropriate dimensions. The real part and imaginary part of $z\in \mathbb{C}$ are denoted by $\rep\rbkt{z}$ and $\imp \rbkt{z}$, respectively. The angle of a scalar $z\in \mathbb{C}$ in the polar form $\abs{z}e^{j\angle z}$ is denoted by $\angle z$. If $z=0$ or $z=\infty$, then $\angle z$ is undefined. 

\subsection{Signals and Systems} 
Throughout, let $\Delta(\cdot)$ denote the difference between two signals labeled by $(\cdot)_1$ and $(\cdot)_2$, e.g., $\Delta u = u_1 - u_2$. Denote the set of energy-bounded $\rn$-valued signals by
\bex
\lt^n \coloneqq \bbkt{u\colon \mathbb{R}_{\geq 0}\to \rn~\bigg| \norm{u}_2^2 \coloneqq \int_{0}^{\infty} |u(t)|^2\,dt < \infty},
\eex 
where the superscript $n$ is dropped when the dimension is clear from the context. For $u, v \in \mathcal{L}_2$, the inner product is denoted as $\ininf{u}{v}= \int_{0}^{\infty} u(t)^\top v(t)\,\mathrm{d} t.$ For $T\geq 0$, define the truncation operator $\trucT$ on a signal $u\colon \mathbb{R}_{\geq 0} \rightarrow \rn$ to be
\begin{equation*}
 (\boldsymbol{\Gamma}_T u)(t)= \left\{ 
 \begin{aligned}
 &u(t)\quad\text{when}~t\leq T\\
 &0\quad\quad\; \text{when}~t> T.
 \end{aligned}
 \right.
\end{equation*}
We often adopt $u_T \coloneqq \boldsymbol{\Gamma}_T u$ for any $T\geq 0$ when there is no ambiguity. Denote the extended space of $\lt$ by 
\bex
\ltep \coloneqq \left\{ u\colon \mathbb{R}_{\geq 0} \to \rn \mid u_T \in \lt~\forall T\geq 0 \right\}.
\eex
For $u, v\in \ltep$ and $T\geq 0$, define the semi-inner product by $\ininf{u}{v}_T \coloneqq \ininf{u_T}{v_T}$ and the semi-norm $\norm{u}_T\coloneqq \norm{ u_T}_2$.

Given two signals $u, v\in \ltp$, define the \emph{phase} \cite{Chen:21_angle, Chaffey:21j} $\theta(u, v)\in \interval{0}{\pi}$ from $u$ to $v$ by
 \be\label{eq: signal_phase}
 \theta(u, v)\coloneqq \arccos \frac{\ininf{u}{v}}{\norm{u}_2\norm{v}_2}
 \ee 
 if $u,v\neq 0 $, and $\theta(u, v)\coloneqq 0$, otherwise. In parallel, define the \emph{gain} $\gamma(u, v)\in \interval{0}{\infty}$ from $u$ to $v$ by
\be\label{eq: signal_gain}
\gamma(u, v)\coloneqq \frac{\norm{v}_2}{\norm{u}_2}
\ee
if $u \neq 0$, and $\gamma(u, v)\coloneqq \infty$, otherwise. For $u, v\in \ltep$ and $T\geq 0$, we have $u_T\in \ltp$ and $v_T\in \ltp$, and similarly define $\theta_T(u, v)\coloneqq \theta(u_T, v_T)~\text{and}~\gamma_T(u, v)\coloneqq \gamma(u_T, v_T)$.

An operator $\sysp\colon \ltep\to\ltep$ is said to be causal if $\boldsymbol{\Gamma}_T\sysp=\boldsymbol{\Gamma}_T\sysp\boldsymbol{\Gamma}_T$ for all $T\geq 0$. We view a system $\sysp$ as a causal operator from input signals to output signals, particularly mapping the zero signal to the zero signal, i.e. $\sysp0=0$. Moreover, only systems with the same number of inputs and outputs are considered. These systems are assumed to be nonzero, i.e. $\sysp\neq 0$. The $\lt$-domain of $\sysp$, i.e. the set of all its input signals in $\ltp$ such that the output signals are in $\ltp$, is denoted by $\mathrm{dom}(\sysp)\coloneqq \left\{ u\in \ltp \mid \sysp u \in \ltp \right\}.$

For causal systems, we adopt the following definition of stability in the finite-gain sense \cite[Section~2.4]{Zames:66}. A causal system $\sysp:\ltep \to \ltep$ is said to be 
\textit{bounded} if $\mathrm{dom}(\sysp) = \ltp$ and 
\bex 
\norm{\sysp} \coloneqq \sup_{0\neq u \in \ltp} \gamma\rbkt{u, \sysp u} < \infty.
\eex 
Moreover, $\sysp$ is said to be \emph{stable} (a.k.a. {continuous}) if it is bounded and 
\bex 
 \norm{\sysp}_{\RI} \coloneqq \sup_{\substack{u_1, u_2\in \ltp\\ u_1\neq u_2}} \gamma\rbkt{u_1- u_2, \sysp u_1 - \sysp u_2} < \infty. 
\eex
System stability is thus regarded as a \emph{continuity} property: small increments in the input lead to small increments in the output. The quantities $\norm{\sysp}$ and $\inorm{\sysp}$ are called the \emph{gain} and the \emph{incremental gain} of the system $\sysp$, respectively. When $\sysp$ is linear, $\norm{\sysp}=\inorm{\sysp}$. Thus, there is no distinction between bounded linear systems and stable linear systems. 

The (topological) \emph{graph} of a causal system $\sysp:\ltep\to\ltep$ is defined by \cite{Georgiou:97}
\bex
\gra(\sysp)\coloneqq \left\{\tbo{u}{\sysp u} \in \ltp\times \ltp \right\}.
\eex
Clearly, here $u$ is restricted to $\dom(\sysp)$ since $\sysp u \in \mathcal{L}_2$. Similarly, the \emph{inverse graph} $\gra^{\dagger}(\sysp)\coloneqq \left\{\stbo{\sysp u}{u} \in \ltp \times \ltp \right\}$ can be defined by swapping the order of inputs and outputs. Likewise, the extended graph and the extended inverse graph of $\sysp$ are defined by \bex \gra_e(\sysp)\coloneqq \left\{\tbo{u}{\sysp u} \in \ltep \times \ltep \right\}\eex and $\gra^{\dagger}_e(\sysp)\coloneqq \left\{\stbo{\sysp u}{u} \in \ltep \times \ltep \right\}$, respectively.

\subsection{The Incremental Directed Gap}
The directed gap \cite{Georgiou:97} between two nonlinear systems provides a measure of how different the two systems' behaviors are. The following definition is adopted. 
\begin{definition}[\hspace{1sp}Directed gap\cite{Georgiou:97}]\label{def: gap}
For two causal systems $\sysp_0, \sysp_1\colon \ltep\to\ltep$, the \textit{directed gap} between $\sysp_0$ and $\sysp_1$ is defined as
\begin{equation}\label{def:gap}
\hspace{-3.1mm}{\delta}(\sysp_0, \sysp_1) \coloneqq \\ \limsup_{T>0} \sup_{\substack{w\in\mathscr{G}_e(\sysp_1),\\\norm{ w_T}_2\neq 0 }} \inf_{\substack{v\in\mathscr{G}_e(\sysp_0),\\ \norm{ v_T}_2\neq 0}} \frac{\norm{ (w - v)_T}_2}{\norm{ v_T}_2}.
\end{equation}
\end{definition}

\vspace{1mm}

In Definition~\ref{def: gap}, we often regard $\sysp_0$ as a \emph{nominal} system and $\sysp_1$ as a \emph{perturbed} system. The value ${\delta}(\sysp_0, \sysp_1)$ quantifies the size of the perturbation from the nominal one $\sysp_0$ to its perturbation $\sysp_1$. Inspired by the directed gap, we propose an incremental counterpart to the directed gap, which differs slightly from the first incremental definition in \cite[p.~90]{Georgiou:95}. 

\begin{definition}[Incremental directed gap]
 For two causal systems $\sysp_0, \sysp_1\colon \ltep\to\ltep$, the \emph{incremental directed gap} between $\sysp_0$ and $\sysp_1$ is defined as 
\begin{multline}\label{def:inc_gap}
{\delta}_{\mathrm{I}}(\sysp_0, \sysp_1) \coloneqq \\
 \limsup_{T>0} \sup_{\substack{w_1, w_2\in\mathscr{G}_e(\sysp_1),\\\norm{ (\Delta w)_T}_2\neq 0 }} \inf_{\substack{v_1, v_2\in\mathscr{G}_e(\sysp_0), \\ \norm{ (\Delta v)_T}_2\neq 0}} {\frac{\norm{ (\Delta w - \Delta v)_T}_2}{\norm{ (\Delta v)_T}_2}}.
\end{multline}
\end{definition}

\vspace{1mm}

The above notion quantifies how different the two systems’ incremental behaviors are, which has a flavor similar to the interpretation of the directed gap. In the case where $\sysp_0$ and $\sysp_1$ are linear systems, we have 
$
{\delta}(\sysp_0, \sysp_1) ={\delta}_{\mathrm{I}}(\sysp_0, \sysp_1).
$

Equipped with ${\delta}(\cdot, \cdot)$ and ${\delta}_{\mathrm{I}}(\cdot, \cdot)$, we introduce a homotopy description for a collection of nonlinear systems parameterized by $\tau \in \interval{0}{1}$. Given a causal system $\sysp: \ltep\to\ltep$, consider a set of causal systems parameterized by $\tau$ 
\be\label{eq: homotopy_set}
\bbkt{\sysp_\tau: \ltep\to\ltep \mid \tau\in \interval{0}{1}, \sysp=\sysp_1}.
\ee
 The set~\eqref{eq: homotopy_set} is shortened as $\sysp_\tau$ for notational brevity. We say $\sysp_\tau$ depends continuously on $\tau \in [0, 1]$ in the directed gap ${\delta}(\cdot)$ if
\bex
{\delta}(\sysp_{\tau}, \sysp_{\nu}) \rightarrow 0~\text{as}~\abs{\nu-\tau}\rightarrow 0
\eex
for all $\tau, \nu \in \interval{0}{1}$.
Analogously, $\sysp_\tau$ is said to depend continuously on $\tau$ in the incremental directed gap ${\delta}_{\mathrm{I}}(\cdot)$ if
\bex
{\delta}_{\mathrm{I}}(\sysp_{\tau}, \sysp_{\nu}) \rightarrow 0~\text{as}~\abs{\nu-\tau}\rightarrow 0
\eex 
for all $\tau, \nu \in \interval{0}{1}$. A commonly adopted homotopy-based characterization is given by
\begin{align}\label{eq: constant_homotopy}
 \sysp_\tau = \{ \tau \sysp \mid\tau \in \interval{0}{1} \}.
 \end{align}

\subsection{Feedback Stability}

 \begin{figure}[htb]
 \centering
 \setlength{\unitlength}{1mm}
 \begin{picture}(50,25)
 \thicklines \put(0,20){\vector(1,0){8}} \put(10,20){\circle{4}}
 \put(12,20){\vector(1,0){8}} \put(20,15){\framebox(10,10){$\sysp$}}
 \put(30,20){\line(1,0){10}} \put(40,20){\vector(0,-1){13}}
 \put(38,5){\vector(-1,0){8}} \put(40,5){\circle{4}}
 \put(50,5){\vector(-1,0){8}} \put(20,0){\framebox(10,10){$\sysc$}}
 \put(20,5){\line(-1,0){10}} \put(10,5){\vector(0,1){13}}
 \put(5,10){\makebox(5,5){$y_2$}} \put(40,10){\makebox(5,5){$y_1$}}
 \put(0,20){\makebox(5,5){$d_1$}} \put(45,0){\makebox(5,5){$d_2$}}
 \put(13,20){\makebox(5,5){$u_1$}} \put(32,0){\makebox(5,5){$u_2$}}
 \end{picture}\caption{A feedback system $\gof$.} \label{fig:feedback}
 \end{figure}

 Consider a positive feedback system illustrated by Fig.~\ref{fig:feedback}, where $\boldsymbol{P}, \boldsymbol{C}\colon \ltep^n \rightarrow \ltep^n$ are causal systems, $d_1$ and $d_2$ are external signals, and $u_1, u_2, y_1$ and $y_2$ are internal signals. These signals are governed by the following equations:
 \bex
 u=d+\tbt{0}{\boldsymbol{I}}{\boldsymbol{I}}{0}y\quad\text{and}\quad y=\tbt{\boldsymbol{P}}{0}{0}{\boldsymbol{C}}u,
 \eex
 where $u=\left[u_1^\top~u_2^\top\right]^\top$, $d=\left[{d_1^\top}~{d_2^\top}\right]^\top$ and $y=\left[{y_1^\top}~{y_2^\top}\right]^\top$. 
 
 For notational brevity, we denote the feedback system by $\gof$, and the associated feedback mapping $u \mapsto d$ by 
 \bex
 \fpc\coloneqq \tbt{\sysi}{-\sysc}{-\sysp}{\sysi}: \ltep^{2n} \rightarrow \ltep^{2n}.
 \eex 
 The input-output (finite-gain) stability of a feedback system $\gof$ is defined as follows.

\begin{definition}[Feedback well-posedness and stability\cite{Zames:66}]\label{def:feedback_stability}
A feedback system $\gof$ is said to be \emph{well-posed} if $\fpc$
has a causal inverse on $\ltep^{2n}$. A well-posed feedback system $\gof$ is said to be \emph{bounded} if $\norm{(\fpc)^{-1}}<\infty$. In addition, $\gof$ is said to be \emph{stable} if it is bounded and $\norm{(\fpc)^{-1}}_{\RI}<\infty$.
\end{definition}

\section{The Scaled Relative Graph of Unbounded Nonlinear Systems}\label{sec: srg}
This section presents the SRG definition of possibly unbounded nonlinear systems and underlines motivation based on an existing SRG separation theorem for feedback stability.

\subsection{SRG Definition}

For a causal system $\sysp\colon \ltep \to \ltep$, based on the gain and phase pairs in \eqref{eq: signal_phase} and \eqref{eq: signal_gain}, we define the \emph{scaled relative graph} \cite{Ryu:21, Chaffey:21j, Chen:25SRG} of $\sysp$ to be
\begin{multline}\label{eq: def_soft_srg}
\hspace{-3mm} \srg(\sysp)= \left\{\gamma(\Delta u, \Delta y) e^{\pm j \theta(\Delta u, \Delta y)} \in \bar{\mathbb{C}}~\Big| \tbo{u_1}{y_1}\in \gra(\sysp), \right. \\
 \left. \vphantom{\tbo{u_2}{y_2}} \tbo{u_2}{y_2} \in \gra(\sysp), \Delta u \neq 0, \Delta y \neq 0 \right\}.
 \end{multline}
 Each pair of input-output incremental trajectories of $\sysp$ corresponds to a complex scalar $z$. The magnitude $\abs{z}$ and argument $\angle z$ contain the incremental gain and incremental phase \cite{Chen:21_angle} information of $\sysp$, respectively. Since $\theta(\cdot, \cdot)$ takes values in $\interval{0}{\pi}$, $\srg(\sysp)$ contains the term $e^{\pm j \theta(\cdot, \cdot)}$, naturally making itself symmetric about the real axis that is consistent with the classical Nyquist plot. We refer the reader to \cite[Proposition~1]{Chaffey:21j} for illustrations of SRGs of some typical bounded nonlinear systems and to \cite[Theorem~4]{Chaffey:21j} for the hyperbolic-convex connection between SRGs and Nyquist plots for stable SISO LTI systems. 

Note that the definition of $\srg(\sysp)$ in \eqref{eq: def_soft_srg} is based on the graph $\gra(\sysp)$ or $u\in \dom(\sysp)$, thereby also encompassing possibly unbounded systems whose $\dom(\sysp)$ being a proper subset of $\lt$. A typical linear example is illustrated below.
 \begin{example}\label{exmp: LTI}
Consider a system $\sysp$ with transfer function $$P(s)=\frac{k}{s},$$
where $k>0$, i.e. a pure integrator. The domain of $\sysp$ can be constructed as $ \dom(\sysp)= \frac{s}{s+p}\mathcal{L}_2$, where $p>0$ can be chosen arbitrary. It is straightforward to verify that \cite{Chen:25SRG}
$$\srg(\sysp) =j\mathbb{R} \setminus \{0\}$$ 
for any $k>0$, which coincides with the Nyquist plot of $P(s)$ without activating the unstable pole at $s=0$.
\end{example}

In addition to Example~\ref{exmp: LTI}, one may wonder about the general connection between SRGs and Nyquist plots for unstable SISO LTI systems. We claim that the connection is similar to the hyperbolic-convex relationship for stable SISO LTI systems; see \cite[Theorem~4]{Chaffey:21j}. The intuition behind this is that $\srg(\sysp)$ adopts admissible $u\in \dom(\sysp)$ only, whereby unstable modes in $\sysp$ cannot be excited by any input. See also a recent discussion in \cite[Section 2]{Krebbekx:25computing}.

For a general unbounded nonlinear system, constructing its $\lt$-domain is often nontrivial. Fortunately, in many applications, one can often focus on a class of systems whose unbounded modes are determined by their linear components only. For example, consider
$$
\sysp = \sysg +\sysh \quad \text{ or }\quad \sysp = \sysg\sysh
$$
with $\sysh$ being unbounded and LTI, and $\sysg$ being bounded and nonlinear. In both cases, $\dom(\sysp)$ is equal to $\dom(\sysh)$ which is easier to obtain via coprime factorization of $\sysh$. Moreover, $\srg(\sysp)$ can also be roughly estimated from $\srg(\sysh)$ and $\srg(\sysg)$ thanks to the SRG interconnection rules \cite{Ryu:21, Chaffey:21j}.

By swapping the role of input and output of a system $\sysp$, we define the \textit{inverse scaled relative graph} of $\sysp$ by
\begin{multline}\label{eq: def_inverse_SRG}
 \hspace{-4mm}\srg^{\dagger}(\sysp) = \left\{\gamma(\Delta y, \Delta u) e^{\pm j \theta(\Delta y, \Delta u)} \in \bar{\mathbb{C}}~\Big|\tbo{y_1}{u_1} \in \gra^{\dagger}(\sysp), 
\right. \\
 \left. \vphantom{\tbo{y_2}{u_2}} \tbo{y_2}{u_2} \in \srg^\dagger(\sysp), \Delta u\neq 0, \Delta y\neq 0 \right\}.
\end{multline}
The purpose of introducing the inverse version is to facilitate the subsequent statement of feedback stability results.

\subsection{Existing SRG Separation Theorem and Motivation}
One of the core achievements in the present SRG theory is a graphical statement of topological graph separation \cite{Teel:96, Georgiou:97} for nonlinear systems. The first SRG-based stability result for one-channel feedback systems was pioneered in \cite[Theorem~2]{Chaffey:21j}, which was subsequently refined and generalized in several directions, e.g. \cite{Chaffey:24, Chen:25SRG, VanDenEijnden:24, Krebbekx:25_nonlinear}. To enable a consistent comparison with the current paper, we adopt the following statement.

\begin{theorem}[{\hspace{1sp}\cite{Chaffey:21j}, \cite[Theorem~2]{Chen:25SRG}}]\label{thm: softSRG}
 Consider a feedback system $\gof$ with $d_2=0$, where $\sysp$ and $\sysc$ are stable. Suppose that $(\tau \sysp)\,\#\, \sysc$ with $d_2=0$ is well-posed for all $\tau \in \interval[open left]{0}{1}$. Then, $\gof$ with $d_2=0$ is stable if
 \be\label{eq: soft_SRG_separation}
 \inf_{\substack{z_{1} \in \srg(\tau \sysp), \tau \in \interval[open left]{0}{1}\\ z_{2} \in \srg^\dagger( \sysc)}} \abs{ z_1 - z_2} >0.
 \ee
 \end{theorem}
\vspace{1mm}

 In Theorem~\ref{thm: softSRG}, given that $\sysp$ and $\sysc$ are both stable, a simple yet powerful homotopy-based characterization as in \eqref{eq: constant_homotopy} is adopted, i.e. $\sysp_\tau = \{ \tau \sysp \mid \tau \in \interval{0}{1} \}$. The SRG composition rules \cite{Chaffey:21j} further facilitate the verification of $\srg(\tau \sysp)$ for $\tau\in \interval[open left]{0}{1}$ as long as $\srg(\sysp)$ is known or can be roughly sketched. Hence, the homotopy-based separation condition in \eqref{eq: soft_SRG_separation} can be viewed as a \emph{mild} technical compromise added to the direct separation of $\srg(\sysp)$ and $\srg^{\dagger}(\sysc)$ in the complex plane. The central pillar of such homotopy-based separation conditions is gap-based robust stability results \cite{Georgiou:97}, which will be elaborated in Appendix.
 
However, when one of the two systems is unbounded, to the best of our knowledge, the above simple homotopy-based description does not work. This can be intuitively demonstrated from two possible cases. First, if $\sysc$ is unbounded, then $(\tau \sysp)\,\#\,\sysc$ when $\tau=0$ results in an unbounded feedback system. This violates the stability assumption of the nominal feedback system when $\tau =0$. Second, if $\sysp$ is unbounded, then $\tau \sysp$ does not depend continuously on $\tau$ in the directed gap or the incremental directed gap. In either case, the homotopy-based argument in \eqref{eq: soft_SRG_separation} is no longer applicable. This somewhat severely hinders the further advancement of the SRG theory. In particular, we even know the exact SRGs of some commonly encountered unbounded systems, as shown in Example~\ref{exmp: LTI}.

One remedy for the above issue is to adopt loop-transformation-based techniques for the original feedback system. Yet, in this paper we aim to address the issue in a more direct fashion. This motivates us to bring more sophisticated homotopy-based descriptions considered in nonlinear robust control \cite{Georgiou:97} into the SRG theory.

\section{Homotopy-Based SRG Separation for Feedback Stability Analysis}\label{sec: main}
 
This section develops the main result, a homotopy-based SRG separation theorem for one-channel feedback stability analysis by exploiting the incremental directed gap. The result allows one of the open-loop systems to be possibly unbounded, while at the price of imposing a generic homotopy-based requirement on the system and and requiring a technical assumption.. We further specialize the result to feedback boundedness analysis and the case of linear systems.

\subsection{SRG Separation}

The following technical assumption will be needed to state the main result. 
\begin{assumption}\label{assumption}
 For a causal system $\sysp\colon \ltep\to \ltep$, suppose that there exists $L>0$ such that
 \be\label{eq: assumption}
 \norm{(\Delta y)_T -(\Delta x)_T}_2 \leq L \norm{ (\Delta u)_T -(\Delta v)_T}_2
 \ee 
 for $T>0$ and $\tbo{u_1}{y_1}, \tbo{u_2}{y_2}, \tbo{v_1}{x_1}, \tbo{v_2}{x_2}\in \gra_e(\sysp)$. 
 \end{assumption}

\vspace{1mm}
Roughly speaking, it is assumed that the system should have a ``second-order incremental gain''. When $\sysp$ is a linear system, Assumption~\ref{assumption} reduces to the boundedness of $\sysp$. The reason for requiring Assumption~\ref{assumption} will be made clear soon. 

Now, we are ready to present the main result. 
\begin{theorem}[SRG separation]\label{thm: SRG_inc_gap}
 Consider a feedback system $\gof$ with $d_2=0$, where $\sysc$ satisfies Assumption~\ref{assumption}. Suppose there exists a mapping $\tau \mapsto \sysp_{\tau}$ in \eqref{eq: homotopy_set} such that 
 \begin{enumerate}
 \renewcommand{\theenumi}{\textup{(\roman{enumi})}}\renewcommand{\labelenumi}{\theenumi}
 \item $\sysp_\tau$ depends continuously on $\tau$ in ${\delta}_{\mathrm{I}}(\cdot, \cdot)$; \label{item: thm_continuous_gap}
 \item $\sysp_0 \,\#\, \sysc$ with $d_2=0$ is stable; \label{item: thm_nominal}
\item $\sysp_\tau\,\#\, \sysc$ with $d_2=0$ is well-posed for all $\tau \in \interval{0}{1}$. \label{item: thm_wellposed}
 \end{enumerate}
If the SRG separation condition
 \be \label{eq:thm_srg_homotopy}
 \inf_{\substack{z_{1} \in \srg(\sysp_\tau ), \tau \in \interval[open left]{0}{1} \\ z_{2} \in \srg^\dagger(\sysc)}} \abs{ z_1 - z_2} >0
 \ee 
 holds, then $\gof$ with $d_2=0$ is stable.
\end{theorem}
\begin{proof}
See Appendix.
\end{proof}

In Theorem~\ref{thm: SRG_inc_gap}, we need to search a homotopy-based description $\sysp_\tau$ that arises from the inclusion of the unbounded system $\sysp$ in the feedback stability analysis. Such a technique has been considered in nonlinear
robust control \cite{Georgiou:97} and the integral quadratic constraint (IQC) literature \cite{Rantzer:97, Khong:21}. The key idea behind the technique is the preservation of robust feedback stability along the entire homotopy path quantified by the incremental directed gap, as elaborated in Theorem~\ref{lem:robust_stability_oneside}\ref{item: robust_stability} in Appendix. Specifically, condition~\ref{item: thm_continuous_gap} assumes the existence of a homotopy path $\sysp_\tau$; condition~\ref{item: thm_nominal} ensures the stability of the nominal feedback system when $\tau =0$; condition~\ref{item: thm_wellposed} is mild, requiring the feedback system to be well-defined along the entire path. Then, the SRG separation condition \eqref{eq:thm_srg_homotopy} required for all $\tau\in \interval{0}{1}$ yields a robust stability margin independent of $\tau$. This enables a continuous perturbation of the feedback system $\sysp_\tau\,\#\, \sysc$ with $d_2=0$ from $\tau = 0$ to $\tau =1$ without loss of stability, measured by the incremental directed gap. Moreover, Theorem~\ref{thm: SRG_inc_gap} still holds if the separation in \eqref{eq:thm_srg_homotopy} is replaced by that of $\srg^\dagger(\sysp_\tau)$ and $\srg(\sysc)$.

By contrast with the fixed choice $\sysp_\tau=\tau \sysp$ in Theorem~\ref{thm: softSRG}, the whole argument above depends on the construction of suitable $\sysp_\tau$ which is partly \emph{heuristic}. An example is provided in Section~\ref{sec: example}. Other commonly adopted choices include
\begin{align*} 
\sysp_\tau = \{ \tau \sysp + (1-\tau) \sysi \mid \tau \in \interval{0}{1} \}.
\end{align*}
In this case, the SRG of $\sysp_\tau$ can be estimated from $\srg(\sysp)$ and $\srg(\sysi) = \{ 1 \}$ using the interconnection rules \cite{Ryu:21, Chaffey:21j}.

Assumption~\ref{assumption} is purely technical for proving Theorem~\ref{thm: SRG_inc_gap}, which arises from utilizing the incremental directed gap to handle a \emph{one-channel} feedback configuration with the second external input $d_2$ set to zero; see Theorem~\ref{lem:robust_stability_oneside}\ref{item: robust_stability} in Appendix. Nevertheless, we anticipate that this assumption can be relaxed to a standard stability requirement on $\sysc$, similar to the boundedness result in Theorem~\ref{thm:softSG_extension} later. We leave this for future work. The restriction imposed by Assumption~\ref{assumption} is sometimes a cause of trouble. The good news is that when $\sysc$ is a linear stable system, the following useful specialization follows immediately.

\begin{theorem}\label{thm: SRG_linear}
 Consider a feedback system $\gof$, where $\sysc$ is linear and stable. Suppose that there exists a mapping $\tau \mapsto \sysp_{\tau}$ in \eqref{eq: homotopy_set} such that
 \begin{enumerate}
 \renewcommand{\theenumi}{\textup{(\roman{enumi})}}\renewcommand{\labelenumi}{\theenumi}
 \item $\sysp_\tau$ depends continuously on $\tau$ in ${\delta}_{\mathrm{I}}(\cdot, \cdot)$; 
 \item $\sysp_0 \,\#\, \sysc$ is stable;
\item $\sysp_\tau\,\#\, \sysc$ is well-posed for all $\tau \in \interval{0}{1}$.
 \end{enumerate}
If condition~\eqref{eq:thm_srg_homotopy} holds, then $\gof$ is stable. 
\end{theorem}
 \begin{proof}
Since $\sysc$ is linear and stable, Assumption~\ref{assumption} is satisfied, and the one-channel feedback stability of $\gof$ with $d_2=0$ is equivalent to the two-channel feedback stability of $\gof$, e.g. see \cite[Section~8]{Jonsson:01} and \cite{Chen:25SRG}. Therefore, we can invoke Theorem~\ref{thm: SRG_inc_gap} and drop the restriction on $d_2=0$, which completes the proof.
 \end{proof}

Theorem~\ref{thm: SRG_linear} is considered more useful and practical than Theorem~\ref{thm: SRG_inc_gap}. An illustrative example is provided in Section~\ref{sec: example} to support its effectiveness. One may further require $\sysp$ to be linear in Theorem~\ref{thm: SRG_linear}, with $\delta_{\mathrm{I}}(\cdot, \cdot)$ being replaced by $\delta(\cdot, \cdot)$.

\subsection{SG Separation}
When feedback boundedness is considered, weaker than feedback stability, Assumption~\ref{assumption} can be relaxed. To elucidate this, we first need the non-incremental version of the SRG for a causal system $\sysp:\ltep\to \ltep$, called the scaled graph (SG) \cite{Chaffey:21j, VanDenEijnden:24}, defined via
\begin{equation*}
 \sg(\sysp)= \left\{\gamma(u, y) e^{\pm j \theta(u, y)} \in \bar{\mathbb{C}}~\Big| \tbo{u}{y}\in \gra(\sysp), u, y \neq 0 \right\}.
\end{equation*} 
The inverse notion $\sg^{\dagger}(\sysp)$ can be defined in a similar manner to \eqref{eq: def_inverse_SRG}. We present the following boundedness result.

\begin{theorem}[SG separation]\label{thm:softSG_extension}
Consider a feedback system $\gof$ with $d_2=0$, where $\sysc$ is stable. Suppose that there exists a mapping $\tau \mapsto \sysp_{\tau}$ in \eqref{eq: homotopy_set} such that
 \begin{enumerate}
 \renewcommand{\theenumi}{\textup{(\roman{enumi})}}\renewcommand{\labelenumi}{\theenumi}
 \item $\sysp_\tau$ depends continuously on $\tau$ in ${\delta}(\cdot, \cdot)$;
 \item $\sysp_0 \,\#\, \sysc$ with $d_2=0$ is bounded;
 \item $\sysp_\tau\,\#\, \sysc$ with $d_2=0$ is well-posed for all $\tau \in \interval{0}{1}$.
 \end{enumerate}
If the SG separation condition
 \be \label{eq:thm_sg_homotopy}
 \inf_{\substack{z_{1} \in \sg(\sysp_\tau ), \tau \in \interval[open left]{0}{1} \\ z_{2} \in \sg^\dagger(\sysc)}} \abs{ z_1 - z_2} >0
 \ee 
 holds, then $\gof$ with $d_2=0$ is bounded.
\end{theorem}
\begin{proof}
See Appendix.
\end{proof}

Notably, the requirement on the incremental gain of $\sysc$ in Theorem~\ref{thm:softSG_extension} arises from addressing the one-channel robust feedback boundedness (see Theorem~\ref{lem:robust_stability_oneside}\ref{item: robust_boundedness}) using the directed gap. This further hints the need for the stronger Assumption~\ref{assumption} on $\sysc$ in the feedback stability result in Theorem~\ref{thm: SRG_inc_gap}.

\subsection{Illustrative Example}\label{sec: example}

We provide an illustrative example below to demonstrate the effectiveness of Theorem~\ref{thm: SRG_inc_gap} or Theorem~\ref{thm: SRG_linear}. Consider a nonlinear system $\sysp:\ltep\to \ltep$ that is decomposed into a parallel interconnection of two subsystems $\sysh$ and $\sysg$, i.e., $$\sysp=\sysh+\sysg,$$ where $\sysh$ is an integrator with transfer function $H(s)=I/{s}$
and $\sysg$ is nonlinear, bounded and incrementally positive\footnote{\label{footnote}A causal bounded system $\sysp:\ltep \to \ltep$ is called incrementally positive if $\ininf{u_1-u_2}{\sysp u_1 - \sysp u_2}\geq 0~\forall u_1, u_2 \in \lt$. It is called strictly incrementally positive if there exist $\delta, \epsilon>0$ such that $\ininf{u_1-u_2}{\sysp u_1 - \sysp u_2}\geq \delta \norm{u_1-u_2}_2^2 + \epsilon \norm{\sysp u_1 - \sysp u_2}_2^2~\forall u_1, u_2 \in \lt$.}. One may interpret the system $\sysp$ as a nonlinear circuit consisting of a linear capacitor $\sysh$ and a nonlinear resistor $\sysg$. Consider a stable linear strictly positive\footref{footnote}~system $\sysc$ with $\delta, \epsilon>0$. In what follows, we show the stability of the feedback system $\sysp\,\#\,(-\sysc)$ in terms of Theorem~\ref{thm: SRG_linear}.

Construct the following mapping $$\sysp_\tau \coloneqq \sysh_\tau + \sysg$$ for $\tau\in \interval{0}{1}$, where $\sysh_\tau$ has transfer function $$H_\tau (s)= \frac{1}{s+1-\tau}I.$$ 
First, note that $\sysp_1= \sysh_1 + \sysg=\sysp$. It can be verified that $\sysp_\tau$ depends continuously on $\tau\in \interval{0}{1}$ in ${\delta}_{\mathrm{I}}(\cdot, \cdot)$ \cite{Georgiou:97}, which fulfills condition~\ref{item: thm_continuous_gap} of Theorem~\ref{thm: SRG_linear}. Second, $\sysp_0$ is bounded and incrementally positive since both $\sysh_0$ and $\sysg$ are so. It follows from \cite[p.~203]{Desoer:75} that the nominal feedback system $\sysp_0\,\#\,(-\sysc)$ is stable, and thereby condition~\ref{item: thm_nominal} is satisfied. Third, one can show that $\srg(\sysp_\tau)\subset \ccp$ for all $\tau\in \interval{0}{1}$ according to the SRG parallel interconnection rule \cite{Chaffey:21j}. By \cite[Proposition~1]{Chen:25SRG}, $\srg^\dagger(-\sysc)\subset \mathcal{D} \subset\mathbb{C}_{<0}$ is contained in a truncated disk sector $\mathcal{D}$ in the open left half-plane. Then, $\srg(\sysp_\tau)$ and $\srg^\dagger(-\sysc)$ are strictly separated for all $\tau\in \interval{0}{1}$, and thus \eqref{eq:thm_srg_homotopy} is satisfied. In light of Theorem~\ref{thm: SRG_linear}, we conclude that $\sysp\,\#\,(-\sysc)$ is stable.

\section{Conclusion}\label{sec:conclusion}
This paper proposed a novel homotopy-based SRG separation theorem for nonlinear feedback stability involving possibly unbounded open-loop systems. The proposed result utilizes the incremental directed gap quantifying one-channel robust
feedback stability along the homotopy path. In doing so, we showed that an important class of unbounded systems, linear integrators, can be incorporated into feedback loops using the SRG-based analysis.

\appendix 
In this appendix, we first develop a technical result concerning robust feedback stability with a single external input (e.g. one-channel stability \cite[Section~5.1]{Brogliato:20}) based on the notion of incremental directed gap. We then establish the proof of Theorem~\ref{thm:softSG_extension} and that of Theorem~\ref{thm: SRG_inc_gap} sequentially. 
\subsection{One-Channel Robust Feedback Stability}
Given a well-posed feedback system $\gof$, define the following parallel projection operator \cite{Doyle:93, Georgiou:97} as 
\bex
\projtp \coloneqq \tbo{d_1}{d_2} \mapsto \tbo{u_1}{y_1} \colon \ltep^{2n} \to \mathscr{G}_e(\sysp).
\eex 
Historically, the directed gap and the parallel projection operator have been used to examine robust boundedness of a feedback loop. The following lemma lies at the core of the gap-based nonlinear robust control \cite{Georgiou:97, Rantzer:97}. 

\begin{lemma}[\hspace{1sp}{\cite[Theorem~3]{Georgiou:97}}]\label{lem:gap}
Consider two well-posed feedback systems $\boldsymbol{P}_0\,\#\,\boldsymbol{C}$ and $\boldsymbol{P}_1\,\#\,\boldsymbol{C}$. Suppose that $\boldsymbol{P}_0\,\#\,\boldsymbol{C}$ is bounded. Then, $\boldsymbol{P}_1\,\#\,\boldsymbol{C}$ is bounded if
 \bex
 {\delta}(\sysp_0, \sysp_1) < \norm{\boldsymbol{\Pi}_{\mathscr{G}_e(\sysp_0)\parallel\mathscr{G}_e^{\dagger}(\sysc)}}^{-1}.
 \eex 
\end{lemma}

\vspace{1mm}
 
 Lemma~\ref{lem:gap} quantifies the robust boundedness of the nominal feedback system $\boldsymbol{P}_0\,\#\,\boldsymbol{C}$. To be specific, the perturbed feedback system $\boldsymbol{P}_1\,\#\,\boldsymbol{C}$ maintain the boundedness provided that the directed gap between the nominal $\sysp_0$ and perturbed $\sysp_1$ is bounded by some gain information obtained from $\boldsymbol{P}_0\,\#\,\boldsymbol{C}$. The term $\norm{\boldsymbol{\Pi}_{\mathscr{G}_e(\sysp_0)\parallel\mathscr{G}_e^{\dagger}(\sysc)}}^{-1}$ is known as the robust boundedness margin \cite{Georgiou:97} of the nominal feedback system $\boldsymbol{P}_0\,\#\,\boldsymbol{C}$.

The feedback systems in Lemma~\ref{lem:gap} involve two external inputs $d_1$ and $d_2$. The feedback systems in Theorems~\ref{thm: SRG_inc_gap} and~\ref{thm:softSG_extension} have one external input being zero, i.e. $d_2=0$. Thus, a version of Lemma~\ref{lem:gap} for both feedback boundedness and stability with $d_2=0$ that fits the aforementioned setting is necessitated. In general, how to extend the SRG stability theorem to two-channel settings is an open problem \cite{Chen:25SRG}. The following technical result tailored for one-channel feedback systems is important for deriving the main result. 

\begin{theorem}[One-channel robust feedback stability]\label{lem:robust_stability_oneside}
Consider two well-posed feedback systems $\boldsymbol{P}_0\,\#\,\boldsymbol{C}$ and $\boldsymbol{P}_1\,\#\,\boldsymbol{C}$ with $d_2=0$. The following statements hold.
\begin{enumerate}
 \renewcommand{\theenumi}{\textup{(\roman{enumi})}}\renewcommand{\labelenumi}{\theenumi}
 \item \label{item: robust_boundedness}
 Suppose that $\sysc$ is stable and $\boldsymbol{P}_0\,\#\,\boldsymbol{C}$ with $d_2=0$ is bounded. Then, $\boldsymbol{P}_1\,\#\,\boldsymbol{C}$ with $d_2=0$ is bounded if
 \bex
 {\delta}(\sysp_0, \sysp_1) < \frac{1}{2+\norm{\sysc}_{\mathrm{I}}}\norm{\boldsymbol{\Pi}_{\mathscr{G}_e(\sysp_0)\parallel\mathscr{G}_e^{\dagger}(\sysc)}\tbt{\sysi}{0}{0}{0}}^{-1}.
 \eex 
 \item \label{item: robust_stability}
 Suppose that $\sysc$ satisfies Assumption~\ref{assumption} with $L>0$ and $\boldsymbol{P}_0\,\#\,\boldsymbol{C}$ with $d_2=0$ is stable. Then, $\boldsymbol{P}_1\,\#\,\boldsymbol{C}$ with $d_2=0$ is stable if
 \bex
 {\delta}_{\mathrm{I}}(\sysp_0, \sysp_1) < \frac{1}{2+L}\norm{\boldsymbol{\Pi}_{\mathscr{G}_e(\sysp_0)\parallel\mathscr{G}_e^{\dagger}(\sysc)}\tbt{\sysi}{0}{0}{0}}_{\mathrm{I}}^{-1}.
 \eex 
\end{enumerate}
\end{theorem}
\vspace{1mm}

\begin{proof}
Our proof is inspired by the approach adopted in the proof of \cite[Theorem~3]{Georgiou:97}. We first show the proof of statement~\ref{item: robust_boundedness} for feedback boundedness. Suppose that 
\be\label{eq: thm1_gap_eq4}
\norm{\boldsymbol{\Pi}_{\mathscr{G}_e(\sysp_0)\parallel\mathscr{G}_e^{\dagger}(\sysc)}\tbt{\sysi}{0}{0}{0}}=\beta
\ee 
and 
\be\label{eq: thm1_gap_eq5}
 {\delta}(\sysp_0, \sysp_1)<\alpha
\ee
with 
\be\label{eq: thm1_gap_small_gain}
\alpha\beta\rbkt{2+\norm{\sysc}_{\mathrm{I}}}<1.
\ee 
By the well-posedness assumption of $\boldsymbol{P}_1\,\#\,\boldsymbol{C}$ with $d_2=0$, we take any $d = \tbo{d_1}{0}\in \mathcal{L}_{2e}$ and decompose it as $$d=m_1-n_1,$$ where $m_1 \in \mathscr{G}_e(\sysp_1)$ and $n_1 \in \mathscr{G}^\dagger_e(\sysc)$. Following the directed gap in \eqref{def:gap} between $\sysp_0$ and $\sysp_1$ and \eqref{eq: thm1_gap_eq5}, we can find $m\in \mathscr{G}_e(\sysp_0)$, possibly depending on $T>0$, such that 
\be\label{eq: thm1_gap_eq1} 
 \norm{ (m_1-m)_T}_2\leq \alpha \norm{ m_T}_2
\ee
for all sufficiently large $T>0$. 
Since the nominal system $\boldsymbol{P}_0\,\#\,\boldsymbol{C}$ with $d_2=0$ is bounded, by its well-posedness, for the same $d_1$, we can find $n\in \mathscr{G}^\dagger_e(\sysc)$ such that $$m-n=\tbo{d_1}{0}.$$ It must hold that
\begin{align*}
\boldsymbol{\Pi}_{\mathscr{G}_e(\sysp_0)\parallel\mathscr{G}_e^{\dagger}(\sysc)}(m-n) &=\boldsymbol{\Pi}_{\mathscr{G}_e(\sysp_0)\parallel\mathscr{G}_e^{\dagger}(\sysc)} \tbt{\sysi}{0}{0}{0} (m-n)\\
&=m 
\end{align*}
such that by \eqref{eq: thm1_gap_eq4}, we have 
\be\label{eq: thm1_gap_eq2} 
 \norm{ m_T}_2\leq \beta \norm{ (m-n)_T}_2
\ee 
for all $T>0$. 
Noting that $n_1 \in \mathscr{G}^\dagger_e(\sysc)$, $n \in \mathscr{G}^\dagger_e(\sysc)$ and by hypothesis $\sysc$ is stable, we then have 
\bex 
\norm{ \obt{\sysi}{0}(n_1-n)_T}_2\leq \norm{\sysc}_{\mathrm{I}}\norm{ \obt{0}{\sysi}(n_1-n)_T}_2
\eex 
for $T>0$. Moreover, it follows from \eqref{eq: thm1_gap_eq1} and $d_2=0$ that 
\begin{align*}
 \norm{ \obt{0}{\sysi} (n_1-n)_T}_2 &= \norm{ \obt{0}{\sysi} (m_1-m)_T}_2 \\
 &\leq \norm{ (m_1-m)_T}_2 \leq \alpha \norm{ m_T}_2
\end{align*} 
 for sufficiently large $T>0$. This gives that 
\begin{align}\label{eq: thm1_gap_eq3} 
 \norm{(n_1-n)_T}_2 &\leq \norm{ \obt{\sysi}{0}(n_1-n)_T}_2 \notag\\
 &\quad\, + \norm{ \obt{0}{\sysi}(n_1-n)_T}_2 \notag \\
 &\leq \alpha \rbkt{1+\norm{\sysc}_{\mathrm{I}}} \norm{ m_T}_2
\end{align}
for $T>0$, where the first inequality is due to $\norm{\stbo{a}{b}}_2\leq \norm{a}_2+\norm{b}_2$. On the one hand, we have 
\begin{align*}
\norm{d_T}_2&=\norm{ (m_1-n_1)_T}_2 =\norm{(m_1-m+m-n+n-n_1)_T}_2\\
&\geq \norm{ (m-n)_T}_2 - \norm{ (m_1-m)_T}_2-\norm{ (n_1-n)_T}_2\\
&\geq \frac{1}{\beta} \norm{ m_T}_2 -\alpha(2+\norm{\sysc}_{\mathrm{I}})\norm{ m_T}_2 \\
&= \frac{1-\alpha\beta(2+\norm{\sysc}_{\mathrm{I}})}{\beta}\norm{ m_T}_2
\end{align*}
for sufficiently large $T>0$, where the first inequality uses $
\norm{a+b}_2\geq \norm{a}_2-\norm{b}_2$ twice and the second one is based on \eqref{eq: thm1_gap_eq1}, \eqref{eq: thm1_gap_eq2} and \eqref{eq: thm1_gap_eq3}. On the other hand, by \eqref{eq: thm1_gap_eq1} we have
 \begin{align*}
 \norm{ (m_1)_T}_2 &=\norm{ (m_1-m+m)_T}_2 \\
 &\leq \norm{ (m_1-m)_T}_2+\norm{ m_T}_2 \leq (1+\alpha)\norm{ m_T}_2 
 \end{align*}
 for sufficiently large $T>0$. Combining the two aspects together yields 
 \begin{align*}
 \norm{ (m_1)_T}_2 \leq \frac{(1+\alpha)\beta}{1-\alpha\beta\rbkt{2+\norm{\sysc}_{\mathrm{I}}}} \norm{ (m_1-n)_T}_2 \eqqcolon c_0 \norm{ d_T }_2 
 \end{align*}
 for sufficiently large $T>0$, where the inequality uses \eqref{eq: thm1_gap_eq1} and $c_0>0$ is due to \eqref{eq: thm1_gap_small_gain}. With \cite[Proposition~1.2.3]{Van:17}, we thus conclude that $\boldsymbol{P}_1\,\#\,\boldsymbol{C}$ with $d_2=0$ is bounded. 

It remains to show the proof of statement \ref{item: robust_stability} for feedback stability. For notational brevity and avoiding ambiguity, let ${\Delta}(\cdot)= (\cdot)^a-(\cdot)^b$ denote the increment between two signals labeled by the superscripts $a$ and $b$.
Suppose 
\be\label{eq: thm2_gap_eq4}
\norm{\boldsymbol{\Pi}_{\mathscr{G}_e(\sysp_0)\parallel\mathscr{G}_e^{\dagger}(\sysc)}\tbt{\sysi}{0}{0}{0}}_{\mathrm{I}}=\beta
\ee 
and
\be\label{eq: thm2_gap_eq5}
 {\delta}_{\mathrm{I}}(\sysp_0, \sysp_1)<\alpha
\ee
with 
\be\label{eq: thm2_gap_small_gain}
\alpha\beta\rbkt{2+L}<1.
\ee 
By the well-posedness assumption of $\boldsymbol{P}_1\,\#\,\boldsymbol{C}$ with $d_2=0$, take any $d^a\coloneqq \tbo{d_1^a}{0}\in \mathcal{L}_{2e}$ and $d^b\coloneqq \tbo{d_1^b}{0}\in \mathcal{L}_{2e}$, and decompose $d^a$ and $d^b$ as $$d^a=m_1^a-n_1^a~\text{ and }~d^b=m_1^b-n_1^b,$$ where $m_1^a, m_1^b \in \mathscr{G}_e(\sysp_1)$ and $n_1^a, n_1^b \in \mathscr{G}^\dagger_e(\sysc)$. By using the incremental directed gap \eqref{def:inc_gap} between $\sysp_0$ and $\sysp_1$ and according to \eqref{eq: thm2_gap_eq5}, we can find $m^a, m^b\in \mathscr{G}_e(\sysp_0)$, possibly depending on $T>0$, such that 
 \be\label{eq: thm2_gap_eq1} 
 \norm{ \rbkt{ {\Delta}m_1- { \Delta m}}_T}_2 \leq \alpha \norm{ \rbkt{ {\Delta} m }_T}_2 
 \ee
 for all sufficiently large $T>0$. Since $\boldsymbol{P}_0\,\#\,\boldsymbol{C}$ with $d_2=0$ is stable and by its well-posedness, for the same $d_1^a$ and $d_1^b$, one can find $n^a, n^b \in \mathscr{G}^\dagger_e(\sysc)$ such that $$m^a-n^a=\tbo{d_1^a}{0} \text{ and } m^b-n^b=\tbo{d_1^b}{0}.$$ It must hold that $$\boldsymbol{\Pi}_{\mathscr{G}_e(\sysp_0)\parallel\mathscr{G}_e^{\dagger}(\sysc)} \tbt{\sysi}{0}{0}{0} (m^j-n^j)=m^j$$ for $j\in \{a, b\}$ 
 such that by \eqref{eq: thm2_gap_eq4}, 
 \begin{align}\label{eq: thm2_gap_eq2} 
 \norm{ \rbkt{ {\Delta}m}_T}_2 \leq \beta \norm{ \rbkt{ {\Delta} m - {\Delta} n}_T}_2 
 \end{align} 
 for $T>0$. Noting that $n_1^a$, $n_1^b$, $n^a$ and $n^b$ are all elements in $\mathscr{G}^\dagger_e(\sysc)$ and using the assumption of $\sysc$ satisfying \eqref{eq: assumption} with $L>0$, we then have 
 $$\norm{ \obt{\sysi}{0}\rbkt{ {\Delta}n_1- {\Delta}n}_T}_2\leq L \norm{ \obt{0}{\sysi}\rbkt{ {\Delta}n_1- {\Delta}n}_T}_2$$
 for $T>0$. Moreover, it follows from \eqref{eq: thm2_gap_eq1} and $d_2=0$ that 
 \begin{multline*}
 \norm{ \obt{0}{\sysi} \rbkt{ {\Delta}n_1- {\Delta}n}_T}_2 = \norm{ \obt{0}{\sysi} \rbkt{ {\Delta}m_1- {\Delta}m}_T}_2 \\
 \leq \norm{ \rbkt{ {\Delta}m_1- {\Delta}m}_T}_2 \leq \alpha \norm{ \rbkt{{\Delta}m}_T}_2
 \end{multline*} 
for sufficiently large $T>0$. This gives that 
\begin{multline}\label{eq: thm2_gap_eq3} 
 \norm{ \rbkt{ {\Delta} n_1- {\Delta}n}_T}_2 \leq \norm{ \obt{\sysi}{0}\rbkt{ {\Delta}n_1- {\Delta}n}_T}_2 \\
 + \norm{ \obt{0}{\sysi}\rbkt{ {\Delta}n_1- {\Delta}n}_T}_2 \leq \alpha \rbkt{1+L} \norm{ m_T}_2
\end{multline}
 for $T>0$. On the one hand, we have 
 \begin{align*}
 \norm{ \rbkt{ {\Delta} d}_T}_2 &= \norm{ \sbkt{m_1^{a}-n_1^{a}-(m_1^{b}-n_1^{b})}_T}_2 \notag\\
&= \norm{ \rbkt{ {\Delta}m_1- {\Delta}m + {\Delta}m + {\Delta}n - {\Delta}n - {\Delta}n_1}_T}_2 \notag \\
&\geq \norm{ \rbkt{ {\Delta}m- {\Delta}n}_T}_2 - \norm{ \rbkt{ {\Delta}m_1 - {\Delta}m}_T}_2 \notag \\
& \quad \, -\norm{ \rbkt{ {\Delta} n_1 - {\Delta} n }_T}_2\notag \\
&\geq \frac{1}{\beta} \norm{ \rbkt{ {\Delta} m}_T}_2-\alpha\rbkt{2+L} \norm{ \rbkt{ {\Delta}m}_T}_2
\end{align*}
 for sufficiently large $T>0$, where the last inequality uses \eqref{eq: thm2_gap_eq1}, \eqref{eq: thm2_gap_eq2} and \eqref{eq: thm2_gap_eq3}. On the other hand, from \eqref{eq: thm2_gap_eq1} we have
 \begin{align*}
 \norm{ \rbkt{ {\Delta}m_1}_T}_2 &=\norm{ \rbkt{\Delta m_1- {\Delta}m+ {\Delta}m}_T}_2 \\
&\leq \norm{ \rbkt{{\Delta}m_1- {\Delta}m}_T}_2+\norm{ \rbkt{ {\Delta}m}_T}_2 \\
&\leq (1+\alpha)\norm{ \rbkt{ {\Delta}m}_T }_2 
 \end{align*}
 for sufficiently large $T>0$. It then follows that 
 \begin{align*}
 \norm{ \rbkt{ {\Delta} m_1}_T}_2 &\leq \frac{(1+\alpha)\beta}{1-\alpha\beta\rbkt{2+L}} \norm{ \rbkt{ {\Delta} m_1- {\Delta}n}_T}_2 \\
 &\eqqcolon c_0 \norm{ \rbkt{ {\Delta}d}_T }_2
 \end{align*}
 for sufficiently large $T>0$, where $c_0>0$ is due to \eqref{eq: thm2_gap_small_gain}. Then $\boldsymbol{P}_1\,\#\,\boldsymbol{C}$ with $d_2=0$ is stable.
\end{proof}

\subsection{Proofs of Theorem~\ref{thm:softSG_extension} and Theorem~\ref{thm: SRG_inc_gap}}\label{appendix: thm1_proof}

Equipped with Theorem~\ref{lem:robust_stability_oneside}, we show the proofs of Theorem~\ref{thm:softSG_extension} and Theorem~\ref{thm: SRG_inc_gap}, respectively, as the former is simpler.
 
 \emph{Proof of Theorem~\ref{thm:softSG_extension}}:~We divide the proof into three steps. 

\emph{Step~1:} This step utilizes the SG separation condition~\eqref{eq:thm_sg_homotopy} to obtain a constant $c_0$ that will be needed for the robust boundedness margin later. This step shares the same lines of reasoning as those in Step~1 of the proof of \cite[Theorem~2]{Chen:25SRG} in its non-incremental setting. Hence, this step is omitted for brevity. The major difference lies in that $\sysp_\tau$ is possibly unbounded whereby $\mathrm{dom}(\sysp_\tau)\subset \lt$ is required in the proof. By using \eqref{eq:thm_sg_homotopy}, for a feedback system $\sysp_\tau \,\#\,{\boldsymbol{C}}$ with $d_2=0$, we can similarly establish that for all nonzero $ u_1 \in \mathrm{dom}(\sysp_\tau)$, nonzero $u_2\in \lt$ and $\tau \in \interval{0}{1}$, there exists $c_0\in \interval[open]{0}{\infty}$ such that $$\norm{u}_2\leq c_0 \norm{ \sysf_{\sysp_\tau \# \sysc}u}_2.$$
 
\emph{Step~2:} We show that the boundedness of $\sysp_\tau \,\#\,{\boldsymbol{C}}$ with $d_2=0$ implies the boundedness of $(\boldsymbol{P}_{\tau+\nu})\,\#\,{\boldsymbol{C}}$ with $d_2=0$ for all $\nu\in\interval[open]{0}{\mu}$, where $\mu$ is independent of $\tau$. This can be proved by using Step~1 and statement~\ref{item: robust_boundedness} of Theorem~\ref{lem:robust_stability_oneside}.

Specifically, by hypothesis $\rbkt{\sysf_{\sysp_\tau \# \sysc}}^{-1}$ is bounded on $\ltp$ and by Step~1, we have $\norm{\rbkt{\sysf_{\sysp_\tau \# \sysc}}^{-1}}\leq c_0$. Noting that the following equation holds:
\bex 
 \projtp=(\fpc)^{-1} +\tbt{0}{0}{0}{-\sysi},
\eex 
we then have 
\begin{align*}
&\norm{\boldsymbol{\Pi}_{\mathscr{G}_e(\sysp_\tau)\parallel\mathscr{G}_e^{\dagger}(\sysc)}\tbt{\sysi}{0}{0}{0}}\\
 =&\norm{(\boldsymbol{F}_{\sysp_\tau\# \sysc})^{-1}\tbt{\sysi}{0}{0}{0} +\tbt{0}{0}{0}{-\sysi}\tbt{\sysi}{0}{0}{0} }\\
=& \norm{(\boldsymbol{F}_{\sysp_\tau\# \sysc})^{-1}} \norm{\tbt{\sysi}{0}{0}{0} }\leq c_0.
\end{align*}
By setting the following uniform constant: 
\[
 \eta \coloneqq \frac{1}{c_0(2+\norm{\sysc}_{\mathrm{I}})}>0,
 \]
there exists a sufficiently small $\mu>0$, independent of $\tau$, such that the perturbed system $\sysp_{\tau+\mu}$ satisfies that 
\[
 {\delta}(\sysp_{\tau}, \sysp_{\tau+\mu}) < \eta\] and
 \begin{multline*}
 {\delta}(\sysp_{\tau}, \sysp_{\tau+\mu}) \norm{\boldsymbol{\Pi}_{\mathscr{G}_e(\sysp_\tau)\parallel\mathscr{G}_e^{\dagger}(\sysc)}\tbt{\sysi}{0}{0}{0}} (2+\norm{\sysc}_{\mathrm{I}}) \\
 < \eta c_0 (2+\norm{\sysc}_{\mathrm{I}}) = 1
 \end{multline*}
 and ${\delta}(\sysp_{\tau}, \sysp_{\tau+\nu}) < \eta$ for all $\nu\in \interval[open]{0}{\mu}$ by the continuity of ${\delta}(\cdot, \cdot)$. It then follows from statement~\ref{item: robust_boundedness} of Theorem~\ref{lem:robust_stability_oneside} that $(\boldsymbol{P}_{\tau+\nu})\,\#\,{\boldsymbol{C}}$ with $d_2=0$ is bounded for all $\nu\in\interval[open]{0}{\mu}$.
 
\emph{Step~3:} By hypothesis, $\boldsymbol{P}_\tau \,\#\,\boldsymbol{C}$ with $d_2=0$ is bounded when $\tau=0$. By {Step~2}, $\rbkt{\sysf_{(\sysp_{(0+\nu)})\#\sysc}}^{-1}$ with $d_2=0$ is bounded for any $\nu\in\interval[open]{0}{\mu}$. Applying Step 2 again to ${(\sysp_{(0+\nu)})\#\sysc}$ with $d_2=0$ implies that $\rbkt{\sysf_{(\sysp_{(0+2\nu)})\#\sysc}}^{-1}$ with $d_2=0$ is also bounded for $2\nu<2\mu$. Applying Step~2 iteratively for $k\nu$ with $k=2, 3, \ldots,$ and by mathematical induction, $\rbkt{\sysf_{\sysp_\tau \#\sysc}}^{-1}$ with $d_2=0$ is bounded for all $\tau \in \interval{0}{1}$. Then, we conclude that $\gof$ with ${d_2=0}$ is bounded when $\tau=1$.  \hspace*{\fill}~\QED 
 
\emph{Proof of Theorem~\ref{thm: SRG_inc_gap}}:~The full proof is omitted, since it follows the same lines of reasoning as the non-incremental arguments in the proof of Theorem~\ref{thm:softSG_extension}. We only indicate their major differences. Step~1 follows immediately from the proof of \cite[Theorem~2]{Chen:25SRG} with $\dom(\sysp_\tau)\subset \lt$. In Step~2, the statement on the boundedness is replaced by that of the stability. In addition, $\delta(\cdot, \cdot)$ is replaced by ${\delta}_{\mathrm{I}}(\cdot, \cdot)$, and statement \ref{item: robust_stability} of Theorem~\ref{lem:robust_stability_oneside} is used to set the constant $\eta= \frac{1}{c_0(2+L)}$. Finally, Step 3 uses mathematical induction to prove the stability of $\rbkt{\sysf_{\sysp_\tau \#\sysc}}^{-1}$ from $\tau =0$ to $\tau =1$. \hspace*{\fill}~\QED 
  
\bibliographystyle{ieeetr}
\bibliography{SRG}

\end{document}